\documentclass{article}
\usepackage[dvips]{graphicx}
\begin{document}
\title{Effective Viscosity and Time Correlation for the Kuramoto-Sivashinsky Equation} 
\author{Hidetsugu  Sakaguchi\\
Department of Applied Science for Electronics and Materials \\
Interdisciplinary Graduate School of Engineering Sciences\\
 Kyushu University, Kasuga 816-8580, Japan}
\maketitle

\hspace*{5cm}

A shock-like structure appears in a time-averaged pattern produced by the Kuramoto-Sivashinsky equation and the noisy Burgers equation with fixed boundary conditions. We show that the effective viscosity computed from the width of the 
time-averaged shock structure is consistent with that computed from the time correlation of the fluctuations.  The effective viscosity depends on the lengthscale, although our system size is not sufficiently large to satisfy the asymptotic dynamic scaling law.  We attempt to determine the effective viscosity in a finite size system 
with the projection operator method.
\newpage
\section{Introduction}
The Kuramoto-Sivashinsky equation is one of the simplest partial differential equations exhibiting spatiotemporal chaos.\cite{rf:1,rf:2}  In one dimension it has the form
\begin{equation}
h_t=-h_{xx}-h_{xxxx}+\frac{1}{2}(h_x)^2,
\end{equation}
where $h=h(x,t)$ is a real function, $x\in [0,L]$, and the subscripts stand for  partial derivatives.  An equivalent equation is obtained for $u=h_x$ as 
\begin{equation}
u_t=-u_{xx}-u_{xxxx}+uu_{x}.
\end{equation}
It is conjectured that the 
large-scale properties of this equation are described by the noisy Burgers equation.\cite{rf:3,rf:4,rf:5}  
\begin{equation}
u_t=\nu_k u_{xx}+\lambda uu_x+\xi_x(x,t),
\end{equation}
where $\nu_k>0$  represents the effective viscosity on a lengthscale of $1/k$ (where $k$ is the wavenumber), $\xi_x$ is the spatial derivative of the random noise $\xi(x,t)$, and the parameter $\lambda$ is 1 to satisfy  Galilean invariance. 
The random noise is assumed to satisfy $\langle \xi(x,t)\xi(x^{\prime},t^{\prime})\rangle=2D_k\delta(x-x^{\prime})\delta(t-t^{\prime})$. 
The noisy Burgers equation is equivalent to the Karder-Parisi-Zhang equation  for a growing interface with fluctuations.\cite{rf:6}
The dynamic scaling of the noisy Burgers equation (the KPZ equation) 
is expressed as
\[
\langle u_k(\omega)u_k^*(\omega)\rangle=k^{-3/2}g(\omega/k^{3/2})
\]
for small wavenumber $k$ and small frequency $\omega$.
The dynamic scaling of the linear Langevin equation
\begin{equation}
u_t=\nu_k u_{xx}+\xi_x(x,t)
\end{equation} 
for $\nu_k={\rm const}$ and $D_k={\rm const}$  
is described by $\langle u_k(\omega)u_k^*(\omega)\rangle=k^{-2}g(\omega/k^2)$, 
which differs from the behavior of the noisy Burgers equation.
If the effective viscosity and the effective noise strength behave as $\nu(k)\sim k^{-1/2}, D(k)\sim k^{-1/2}$, 
the power spectrum takes the form $k^{-3/2}/\{(\omega/k^{3/2})^2+1\}$, which has the same form as   
the dynamic scaling for the noisy Burgers equation. 

The dynamic scaling regime of the noisy Burgers equation has not been clearly found  in direct numerical simulations of the Kuramoto-Sivashinsky equation. 
Sneppen et al. performed a largescale numerical simulation of the Kuramoto-Sivashinsky equation of the form (1).\cite{rf:7}  They  determined the effective 
diffusion constant as $\nu\sim 10.5$ and the noise strength as $D\sim 6.4$ from analysis of the interface width 
$\langle(h(x,t)-\langle h\rangle)^2\rangle$ in the dynamic scaling regime of the linear Langevin equation.    
They found a crossover toward the dynamic scaling regime of the noisy Burger equation. However, they could not demonstarate the  dynamic scaling itself.
It is believed that a much larger scale numerical simulation is necessary to find the dynamic scaling regime of the noisy Burgers equation.

In this paper, we show that the effective viscosity can be evaluated from the relation between  the width and amplitude of the time-averaged shock structure and the time correlation of the Fourier modes. 
We do not attempt to find the asymptotic dynamic scaling. However, we find 
some dependence of the effective viscosity on the lengthscale.
We also performed similar numerical simulations for the noisy Burgers equation. We attempt to determine the effective viscosity for the Kuramoto-Sivashinsky equation with the projection operator method.

\section{Shock structure in the time-averaged pattern}
We numerically studied the Kuramoto-Sivashinsky equation with  
fixed boundary conditions $u(0)=-U$ and $u(L)=U$ in a previous paper.
\cite{rf:8} The numerical simulations were performed using the Heun method with time step $\Delta t=0.002$ and the space step $\Delta x=0.25$.
Shock-like structures appear in time-averaged patterns 
for a certain range of boundary values.  
A nonlinear effect is essential for the formation of  shock 
structure.  A nonlinear effect appears in this type of simulation even for a small system.    Contrastly,  very large systemsize is necessary to find a nonliner effect in the analysis of a fluctuating interface width.  
The shock structures are approximated by the functional form 
\[u(x)=A\tanh(\kappa \{x-L/2)\}.\]
\begin{figure}
\begin{center}
\includegraphics[width=7cm]{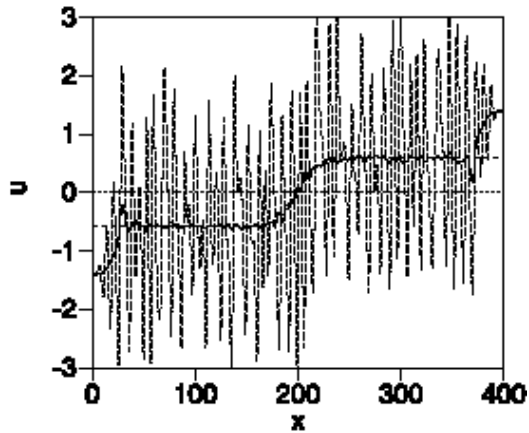}
\end{center}
\caption{Time-averaged shock pattern and a snapshot pattern for the Kuramoto-Sivashinsky equation with  $U=1.4$.}
\label{fig:1}
\end{figure}
Figure 1 displays an example of the shock-like structure for $U=1.4$.
The dashed curve represents a snapshot pattern of $u(x,t)$, the solid curve 
is a time-averaged pattern, and the dotted curve represents $0.58\tanh\{0.055(x-L/2)\}$.  It is seen that the amplitude $A$ of the time-averaged shock structure is not  the boundary value $U$.
The effective viscosity $\nu_\kappa$ is evaluated from the relation $\nu_\kappa=A/(2\kappa)$, where $\lambda=1$ is assumed, since the space step $\Delta x$ is 
sufficiently small. (Sneppen et al. estimated $\lambda=2.37$ for $\Delta x=1$ and $\lambda=1.025$ for $\Delta x=0.5$. We believe that the value of $\lambda$ that they found differs from 1, because the space step $\Delta x=1$ used in their numerical simulation is too large and as a result Galilean invariance is not satisfied.)  

We showed that the effective viscosity  depends on the lengthscale of the shock width $1/\kappa$. The effective viscosity tends to increase as the shock width is increased, and it seems to approach a value near 10 in a large lengthscale. 
The value of the effective viscosity is consistent with the value obtained by Sneppen et al.\cite{rf:6}  

We numerically confirmed that such a time-averaged shock structure appears 
in the noisy Burgers equation (3) with constant bare viscosity $\nu_k=\nu$.  The boundary conditions were fixed as $u(0)=-U$ and $u(L)=U$. The numerical simulation was performed using the Heun method with time step $\Delta t=0.001$ and the space step $\Delta x=0.25$. 
To maintain $\int_0^L u(x)dx=0$, as the case of the Kuramoto-Sivashinsky equation, we have further assumed that $\int_0^L \xi_xdx=0$.
The parameter values are $L=400,\,\nu=10$ and $D=6.5$. The viscosity and the noise strength are chosen to be close to the values found by Sneppen et al.
\begin{figure}
\begin{center}
\includegraphics[width=12cm]{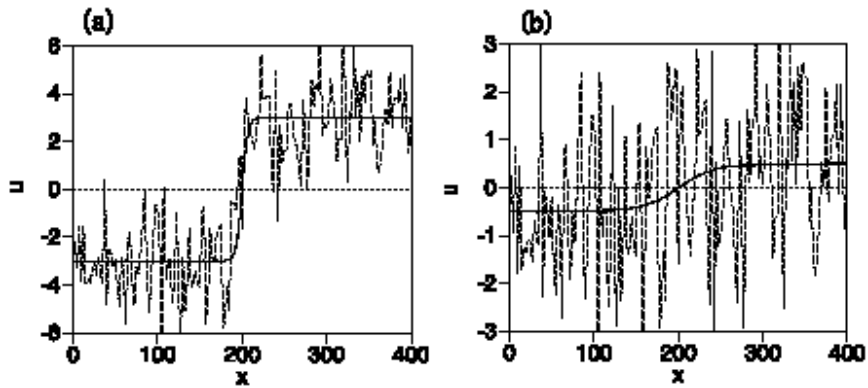}
\end{center}
\caption{Time-averaged shock patterns and snapshot patterns for the nosiy Burgers equation with  $\nu=10$, $D=6.5$ and (a) $U=3$, (b) $U=0.5$.}
\label{fig:2}
\end{figure}
Figure 2 (a) displays a snapshot pattern of $u(x,t)$ (dashed curve), the time-averaged pattern of $\bar{u}(x,t)=(1/T)\int_0^Tu(x,t)dt$ with $T=50000$ (solid curve) for $U=3$, and a plot of the function $\tanh\{0.15(x-L/2)\}$.   
Figure 2 (b) displays a snapshot pattern of $u(x,t)$ (dashed curve), the time-averaged pattern (solid curve) for $U=0.5$, and a plot of the function $\tanh\{0.0227(x-L/2)\}$.   The values of $u(x)$ at $x=x_i=2.5\times i$ with integer $i$  are plotted for the snapshot patterns to show the three plots clearly. 
The time-averaged shock structures appear clearly also for the noisy Burgers equation. It is seen that the amplitude of the time-averaged shock structure is equal to the boundary value $U$. The effective viscosity can be evaluated from the relation between the shock amplitude and the width. The effective viscosities were determined for several values of $U$, and the results are $\nu_{\kappa}=10$ for $\kappa=0.15\, (U=3)$, $\nu_{\kappa}=10$ for $\kappa=0.05\, (U=1)$, $\nu_{\kappa}=11$ for $\kappa=0.0227\, (U=0.5)$, and $\nu_{\kappa}=11.5$ for $\kappa=0.013\, (U=0.3)$.   
These effective viscosities are close to the original viscosity 10. However, they tend to increase slightly as $\kappa$ is decreased.

\section{Time correlation of Fourier modes}
It is conjectured that the large scale properties of the Kuramoto-Sivashinsky equation are described by the noisy Burgers equation. It is not easy to 
evaluate the effect of the nonlinear term even for the noisy Burgers equation. 
We study the statistical properties of the Kuramoto-Sivashinsky equation by reference to  the linear Langevin equation (4) with scale-dependent 
coefficients.  
The Fourier amplitude $u_k(t)$ for the wavenumber $k$ for Eq.~(4) obeys 
\begin{equation}
\frac{d u_k}{dt}=-\nu_k k^2u_k+\xi_k(t),
\end{equation}
where $\xi_k(t)$ is the Fourier transform of $\xi(x,t)$. 
The noise $\xi_k(t)$ is assumed to satisfy 
\[\langle \xi_k(t)\rangle=0,\;\;\langle \xi_k(t)\xi_k(t^{\prime})\rangle=2D_kk^2\delta(t-t^{\prime}).\] 
We further assume that $\nu_k$ and $D_k$ may depend on $k$.

The stationary probability distribution of $u_k$ is Gaussian distribution 
\begin{equation}
P(u_k)\propto \exp\{-u_k^2/(2\sigma_k)\},
\end{equation}
where $\sigma_k=D_kk^2/\{\nu(k) k^2\}$, and the time correlation function is 
\[\langle u_k(t)u_k^*(t^{\prime})\rangle=\sigma_k\exp(-\nu_k k^2 |t-t^{\prime}|).\]
The decay constant is $\tau=1/\{\nu_k k^2\}$. We can estimate the effective viscosity $\nu_k$ from the decay constant as $\nu(k)=1/(\tau k^2)$.
The power spectrum of $u_k$ is expressed as $\langle |u_k(\omega)|^2\rangle\sim D_kk^2/\{\omega^2+\nu_k^2k^4\}$.
\begin{figure}
\includegraphics[width=12cm]{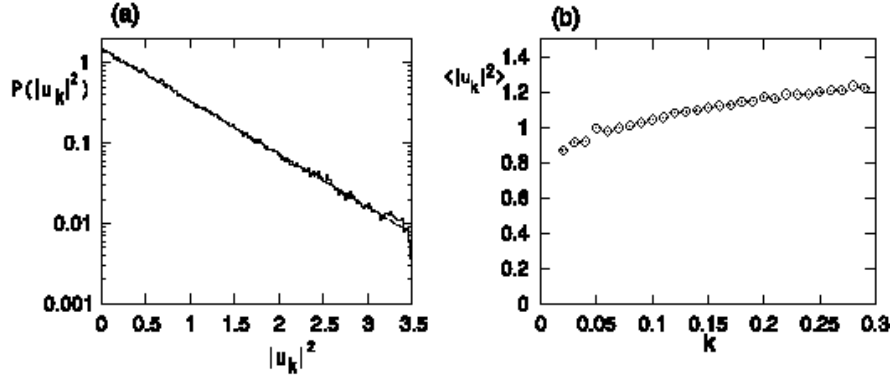}
\caption{(a)Probability distribution $P(|u_k|^2)$ for $k=0.12$ (histogram) and the 
Gaussian approximation $1/\sqrt{2\pi\sigma_k}\exp\{-u_k^2/(2\sigma_k)\}$ (dashed curve) in a semi-logarithmic scale. (b) Energy spectrum $\langle|u_k|^2\rangle$ as a function of $k$.}
\label{fig:3}
\end{figure}

We  numerically studied the statistical properties of the 
temporal 
fluctuation of the Fourier amplitude $u_k(t)$ with small 
wavenumber $k$ for the Kuramoto-Sivashinsky equation (2)  with 
periodic boundary conditions.  
We performed numerical simulations of the Kuramoto-Sivashinsky 
equation (2) using the pseudo-spectral method with 1024 and 2048 modes. 
We used system sizes $L=200\pi$ and $400\pi$, and  
the timestep 0.01. 
Figure 3(a) displays the probability 
distribution $P(|u_k|^2)$ for $k=0.12$, and it is compared with Gaussian distribution $P(|u_k|^2)=1/\sqrt{2\pi\sigma_k}\exp\{-|u_k|^2/(2\sigma_k)\}$, where the numerically obtained value of $\langle|u_k|^2\rangle$ is used for $\sigma_k$. The stationary distribution of the temporal fluctuation $u_k$ can be approximated by the Gaussian distribution.  This is consistent with the statistical properties of the linear Langevin model.  The energy spectrum $\langle|u_k|^2\rangle\sim \sigma_k$ has a peak near $k\sim 0.7$, which corresponds to the characteristic cellular structures in KS turbulence. The energy spectrum is a slightly increasing function of $k$ in the small-wavenumber region, as shown in Fig.~3(b).  This is interpreted as meaning that $D_k/\nu_k$ is a slightly increasing function of $k$ in the small-wavenumber region. 

The time correlation $\langle u_k(t)u_k^*(t^{\prime})\rangle$ 
was numerically calculated for small $k$. 
The time correlation function can be approximated by $\exp(-t/\tau)$, except for  fairly small $t$.  The decay constant $\tau_k$ was numerically evaluated as the time where the normalized correlation $\langle u_k(0)u_k^*(t)\rangle/\sigma_k$ becomes $1/e$.  The effective viscosity $\nu$ was evaluated 
from $\nu_k=1/(\tau_kk^2)$. 
\begin{figure}
\begin{center}
\includegraphics[width=7cm]{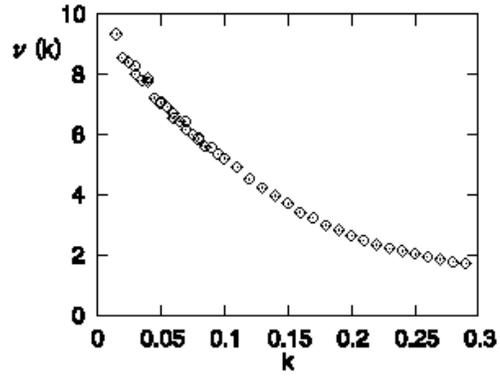}
\end{center}
\caption{Effective viscosity $\nu(k)$ as a function of $k$ for the Kuramoto-Sivashinsky equation.}
\label{fig:4}
\end{figure}
Figure 4 displays values of $\nu_k$ as a function of $k$. 
We see that the effective viscosity is a decreasing function of $k$. 
It approaches a value near 10 as $k$ decreases to 0.  The effective viscosity 
determined  from the width of the shock structure 
was found to be a decreasing function of $\kappa$, where $\kappa$ is the inverse of the shock width.  It approaches a value near 10 for small $\kappa$,  and $\nu_{\kappa}$ is about 5 at $\kappa=0.1$. This behavior is consistent with the present result, although the lengthscales $1/k$ and $1/\kappa$ are slightly different quantities. 

We have also calculated the effective viscosity $\nu_k$ from the time correlation  for the noisy Burgers equation (3) with $\nu=10, D=6.5$ and $L=1024$  and with periodic boundary conditions. 
\begin{figure}
\begin{center}
\includegraphics[width=7cm]{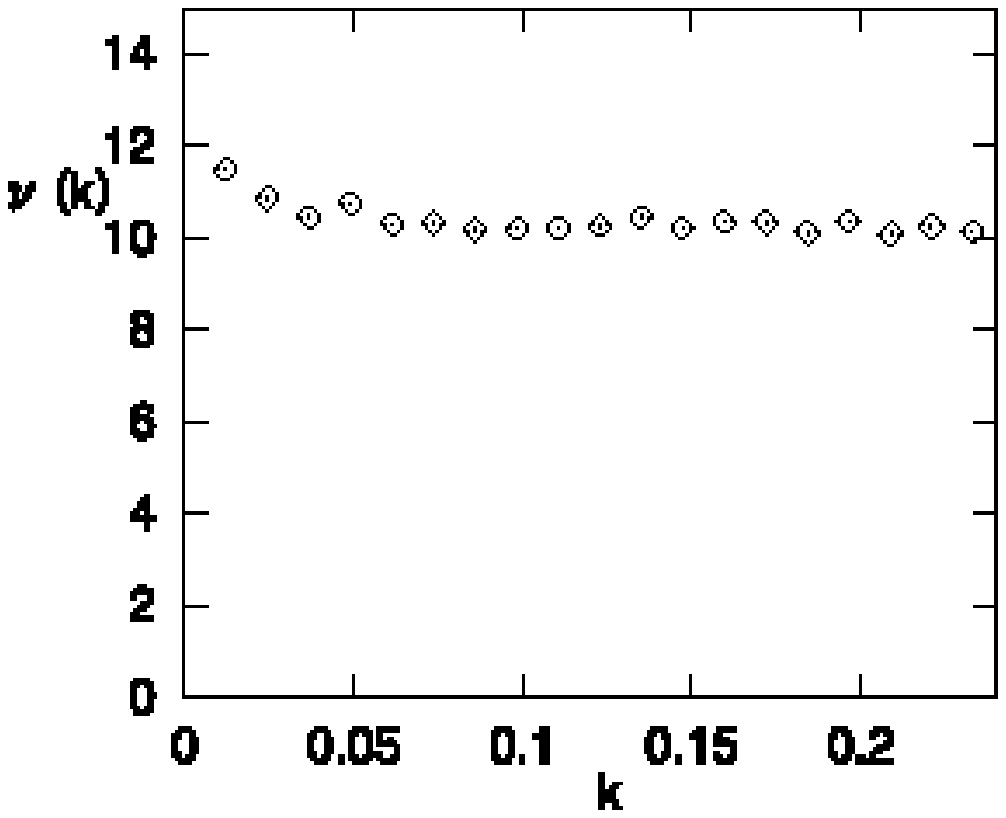}
\end{center}
\caption{Effective viscosity $\nu_k$ as a function of $k$ for the noisy Burgers equation.}
\label{fig:5}
\end{figure}
Figure 5 displays the numerically evaluated effective viscosity $\nu_k$.
We found that the effective viscosity is almost the same as the original viscosity, 10. 
However, the effective viscosities for small $k$ take slightly larger values than the original viscosity. 
The effective viscosity is expected to increase as $k^{-1/2}$ as $k\rightarrow 0$ in the aysmptotic regime of the KPZ dynamic scaling. However, this dependence is not clearly seen in this plot.  It is expected that a much larger scale numerical simulation is necessary to study the asymptotic dynamic scaling regime, even for the noisy Burgers equation itself.   (Sneppen et al. determined a critical lengthscale of the crossover from the dynamic scaling of the linear Langevin equation to that of the noisy Burgers equation as 2500. However, they  could not clearly find the aymptotic dynamic scaling regime in their numerical simulation of system size $65,536$ for the Kuramoto-Sivashinsky equation.  It is believed that a very much larger system size may be necessary, and we have not yet tried such a larger size simulation.)

\section{Evaluation of the effective viscosity using the projection operator method}
We studied statistical properties of the Kuramoto-SIvashinsky equation in the previous section. We now attempt to derive the effective viscosity theoretically for a finite size system. 
Recently, Mori and Fujisaka applied a projection operator method, which was originally developed for a system near thermal equilibrium, to chaotic and turbulent systems.\cite{rf:9}  We now apply their method to the Kuramoto-Sivashinsky equation.

The Kuramoto-Sivashinsky equation can be rewritten as 
\begin{equation}
\frac{du_k}{dt}=(k^2-k^4)u_k+v_k,
\end{equation}
where $v_k=ik/2\sum u_lu_{k-l}$ is the nonlinear term. A projection operator $P$ is defined as 
\begin{equation}
P f(\{u_k\})=\langle f({u_k})u_k^*\rangle/\langle u_ku_k^*\rangle u_k,
\end{equation}
where $f(\{u_k\})$ is a function of $\{u_k\}$, and the average $\langle \cdots\rangle$ is taken with respect to the stationary  distribution of ${u_k}$.  The operators $Q$ and $\Lambda$ are defined as $Q=1-P$ and 
\[\Lambda=\sum(\{(k^2-k^4)u_k+v_k\}\frac{\partial}{\partial u_k}.\]
The operator $\Lambda$ is interpreted as the Liouville operator.
Applying the projection operator method, the Kuramoto-Sivashinsky equation leads to
\begin{equation}
\frac{du_k(t)}{dt}=(k^2-k^4)u_k(t)-\int_0^t\Gamma_k(t-s)u_k(s)ds+r_k(t).
\end{equation}
This equation has the form of a  Langevin equation with the memory term $\int_0^t\Gamma_k(t-s)u_k(s)ds$ and the noise term $r_k(t)$, 
where $r_k(t)$ and $\Gamma(s)$ are expressed as
\[r_k(t)=e^{tQ\Lambda}Qv_k,\]
\[\Gamma_k(s)=\langle r_k(s)r_k^*(0)\rangle/\langle u_ku_k^*\rangle.\]
The above equations are interpreted as constituting the fluctuation dissipation theorem of the second kind. 
The nonlinearity is included in the equation through the terms of $\Gamma_k(s)$ and $r_k(t)$.
To evaluate the effective viscosity, we have approximated 
the effective noise term $r_k(t)$ as $r_k(t)=r_k(0)+tdr_k(0)/dt=Qv_k(0)+tQ\Lambda Qv_k(0)$. 
The effective viscosity of a two-dimensional fluid in thermal equilibrium 
was calculated by Iwayama and Okamoto using a similar type of expansion.\cite{rf:10}

We have further 
approximated the integral $\int_0^t\Gamma_k(t-s)u_k(s)ds$ as $\int_0^t\Gamma_k(s)dsu_k(t)$. This is a kind of Markov approximation.  
Then, the effective damping constant $\gamma_k$ for $u_k$ is approximated as
\begin{equation}
\gamma_k=-k^2+k^4+\int_0^t \langle r_k(s)r_k^*(0)\rangle ds/\langle u_ku_k^*\rangle.
\end{equation}
To calculate $\gamma_k$, we have further assumed that \[\int_0^t\langle r_k(s)r_k^*(0)\rangle ds=\int_0^t\{\langle r_k(0)r_k^*(0)\rangle-s\langle -Q\Lambda Qv_k(0)r_k^*(0)\rangle\}ds=\frac{\langle r_k(0)r_k^*(0)\rangle^2}{2\langle -Q\Lambda Qv_k(0)r_k^*(0)\rangle},\]
where the intergal is approximated at the area of the triangular region satisfying  $\langle r_k(0)r_k^*(0)-s\langle-Q\Lambda Qv_k(0)r_k^*(0)\rangle \,>\,0$.
The effective damping coefficient is expressed as  a complicated 
summation of equal-time correlation functions such as $\langle r_k(0)r_k(0)^*\rangle=\langle Qv_k(0)(Qv_k(0))^*\rangle$, although we have used many rough approximations.
The effective viscosity is evaluated as $\nu_k=\gamma_k/k^2$

\begin{figure}
\begin{center}
\includegraphics[width=7cm]{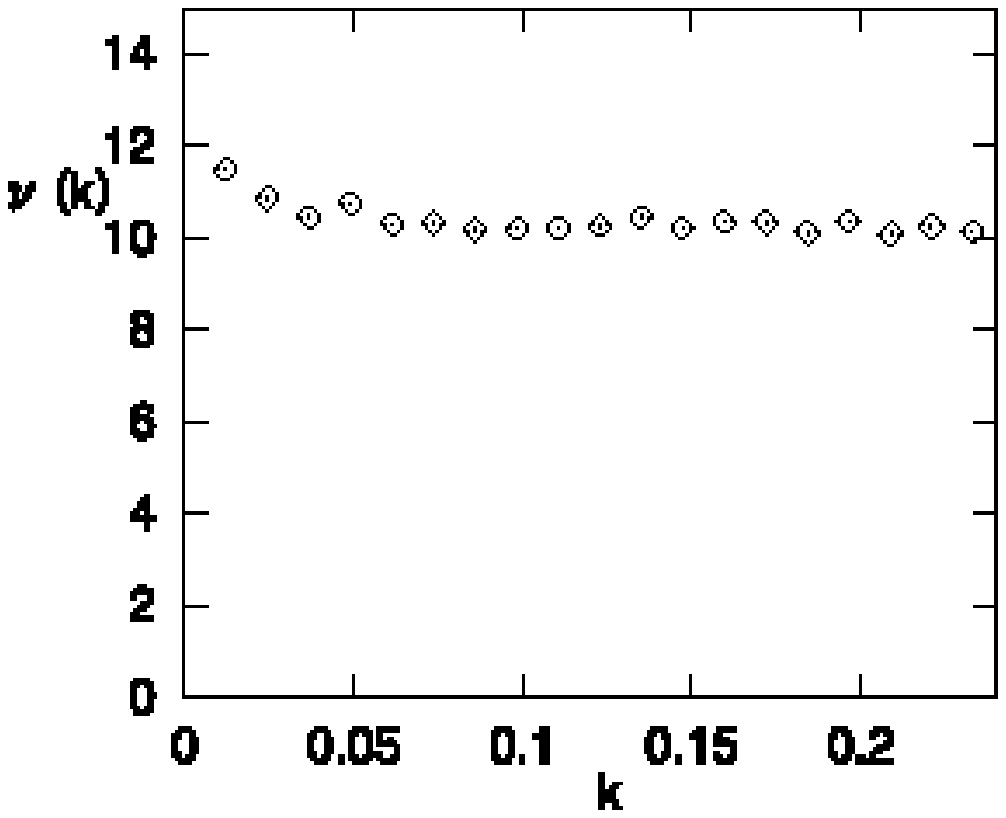}
\end{center}
\caption{Effective viscosity $\nu_k$ as a function of $k$ for the Kuramoto-Sivashinsky equation using the projection operator method.}
\label{fig:6}
\end{figure}
We have computed the effective viscosity $\nu_k$ with a numerical simulation, where the statistical average $\langle\cdots\rangle$ is replaced by the long time average of the chaotic time evolution. Figure 6 displays the numerical results for the system size $L=200\pi$ using the pseudo-spectral method with 1024 modes. 
We obtained positive values for the  effective viscosity and found that it tends to increase as $k$ decreases. However, the effective viscosities found in this way take much larger values than the values obtained directly from the time correlation function in Fig.~4. This might be due to the crudeness of the approximation. For example, it might be necessary to expand to the next order of $t$, although  more complicated summations appear in that case.  However, 
the most essential approximation is the Markov approximation, in which it is assumed that the time correlation of the noise terms is very short. However, the noise term $r_k$ for the wavenumebr $k$ includes the contribution from the slower variables $u_{k^{\prime}}$s with $k^{\prime}<k$, and hence the approximation of the short time correlation  cannot be generally justified.

\section{Summary and discussion}
We have numerically studied the statistical properties of the Kuramoto-Sivashinsky equation. We  calculated the time-averaged pattern with fixed boundary conditions and the time correlation of the temporal fluctuation of the Fourier amplitude with periodic boundary conditions. 
The effective viscosity $\nu_k$ was evaluated from the decay constant of the time correlation function. The numerically obtained value of $\nu_k$ is consistent with the effective viscosity $\nu_{\kappa}$ determined from the shock structure, which is due to stepwise fixed boundary  conditions.  It is shown that the effective viscosity depends on the lengthscale.  It is interpreted as a kind of the fluctuation-dissipation relation for  spatio-temporally chaotic systems that the response  to the external conditions is closely related to the temporal fluctuation in the stationary state.  Both the response function and the temporal fluctuation are described through the effective viscosity.    

The large-scale properties of the Kuramoto-Sivashinsky equation are conjectured to be described by the noisy Burgers equation.  
We have obtained a time-averaged shock structure and the effective viscosity for the noisy Burgers equation. It is found that there is some correspondence between the two equations  even in a finite-size system.

We have attempted to evaluate the effective viscosity for the Kuramoto-Sivashinsky equation using the projection operator method. 
A Langevin-type equation was formally obtained using the projection operator method.  The effective viscosity is formally described in terms of the correlation function of  noise terms. 
We have obtained positive  values of the effective viscosity. However, the values we obtained are larger than the effective viscosity evaluated from the time correlation and the shock width.  We need to improve the approximation.

We have not yet found the dynamic scaling regime of the noisy Burgers equation in numerical simulations of the Kuramoto-Sivashinsky equation, 
although we have found that the effective viscosity tends to increase as $k$ is decreased. 
We need to study the dynamic scaling regime with  numerical simulations of much larger systems.
\section*{Acknowledgement}
We would like to thank Professor M.~Okamura and Mr.~Y.~Kitahara for valuable discussions.  They have studied the time-averaged shock structure in the Kuramoto-Sivashinsky equation directly using the projection operator method.

\end{document}